\documentclass[onecollarge,natbib]{svjour2}
\bibpunct{[}{]}{;}{n}{}{,} 

\usepackage{graphicx}
\usepackage{mathptmx}

\usepackage{amsmath}
\usepackage{feynmp}
\usepackage{tikz}
\usepackage{units}
\usepackage{hyperref}
\usepackage{geometry}
\newcommand{\abs}[1]{\lvert#1\rvert} 
\newcommand*{\pra}{Phys Rev A } 
\newcommand*{\prl}{Phys Rev Lett } 
\newcommand*{\plb}{Phys Lett B } 
\newcommand*{\fbs}{Few Body Syst }
\newcommand*{\ppnp}{Prog Part Nucl Phys }
\newcommand*{\physrep}{Phys Rep } 
\newcommand*{\np}{Nat Phys } 

\journalname{Few-Body Systems (EFB23)}

\title{Trimer and Tetramer Bound States in Heteronuclear Systems}
\subtitle{}

\author{Christiane H. Schmickler \and Hans-Werner Hammer \and Emiko Hiyama}

\institute{C.H. Schmickler 
    \at  {Institut f\"ur Kernphysik,
Technische Universit\"at Darmstadt, 64289 Darmstadt, Germany}
   \at {RIKEN Nishina Center, RIKEN, Saitama 351-0198, Japan}\\
      \email{schmickler@theorie.ikp.physik.tu-darmstadt.de} 
    \and H.-W. Hammer 
\at{Institut f\"ur Kernphysik,
Technische Universit\"at Darmstadt, 64289 Darmstadt, Germany}
\at{ExtreMe Matter Institute EMMI, GSI Helmholtzzentrum
f\"ur Schwerionenforschung GmbH, 64291 Darmstadt, Germany}
\and E. Hiyama
\at{RIKEN Nishina Center, RIKEN, Saitama 351-0198, Japan}
}

\date{Received: date / Accepted: date}

\begin{document}
 \maketitle
 \begin{abstract}The Efimov effect in heteronuclear cold atomic systems is experimentally more easily
accessible than the Efimov effect for identical atoms, because of the potentially smaller
scaling factor. We focus on the case of two or three heavy identical bosons and another atom. 
The former case was recently observed in a mixture of $^{133}$Cs and $^6$Li atoms.

We employ the Gaussian Expansion Method as developed by Hiyama, Kino et al.~\cite{hiyamakino2003}. 
This is a variational method that uses Gaussians that are distributed geometrically
over a chosen range. Supplemental calculations are performed using the Skorniakov-Ter-%
Martirosian equation.

Blume et al.~\cite{blumeyan2014} previously investigated the scaling properties of heteronuclear systems
in the unitary limit and at the three-body breakup threshold. We have completed this
picture by calculating the behaviour on the positive scattering length side of the Efimov
plot, focussing on the dimer threshold. 

 \end{abstract}

  \section{Introduction}
  \label{intro}
\begin{figure}
   \centering
   \begin{tikzpicture}[x=0.9cm,y=0.9cm]
    \node at (1.7,0.5) {\includegraphics[width=0.45\textwidth]{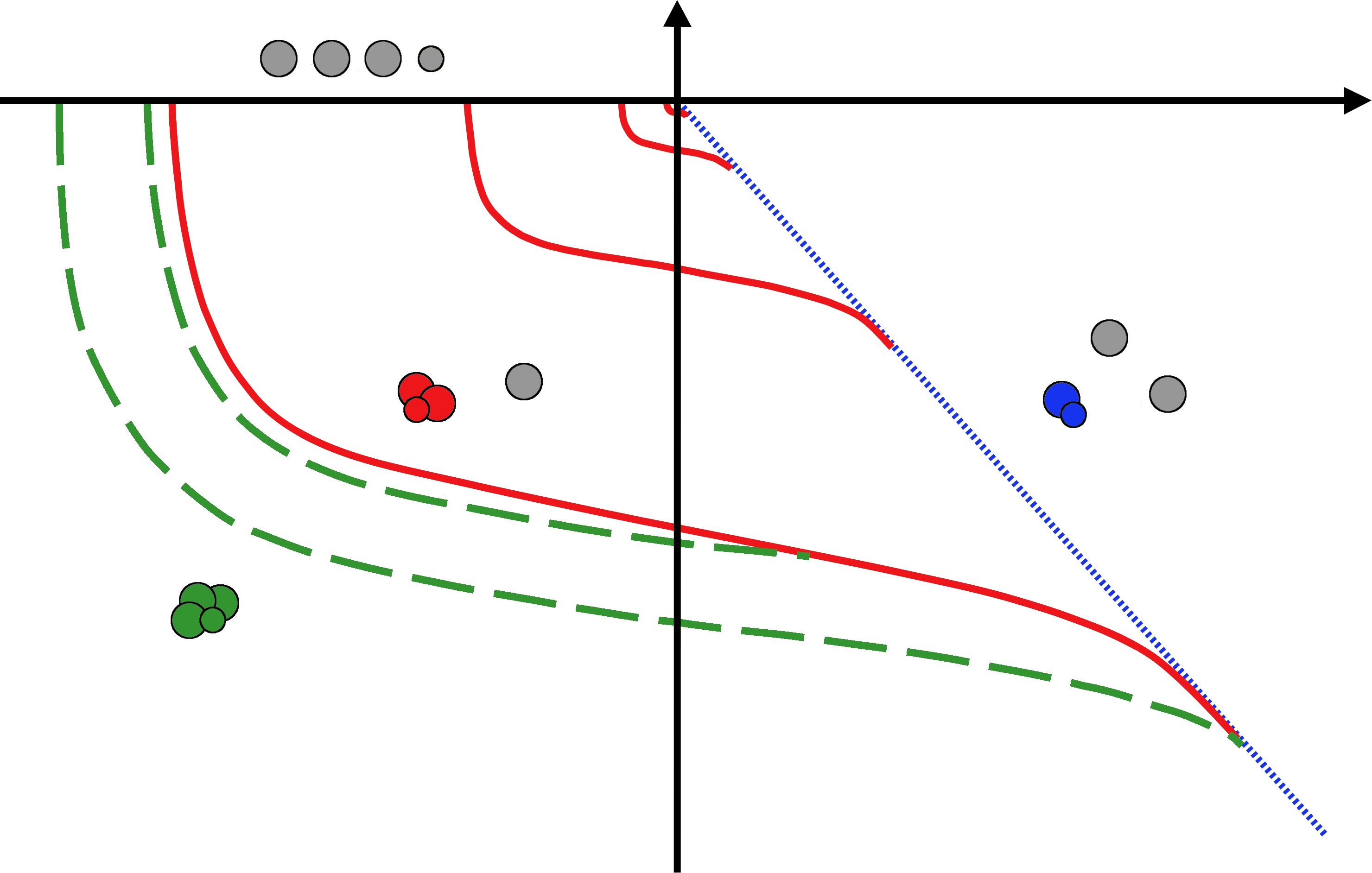}};
    \node at (1.3,3) {$\sqrt{E}$};
    \node at (5.6,2.8){$\frac{1}{a}$};
    \end{tikzpicture}
   \caption{Schematic Efimov plot of the mass imbalanced four-body system. The $H_2L$ trimers are 
   the solid red lines, the $H_3L$ tetramers are the dashed green lines and the $HL$ dimer threshold
   is the dotted blue line. Since we assume no interaction between the $H$ atoms, there is only one 
   dimer.}
   \label{efimovplot}
  \end{figure}

  The Efimov effect has been intensively investigated in recent years, both 
  theoretically and experimentally. It was predicted by Efimov~\cite{efimov1970} that in a system 
  with a scattering length $a$ that is large compared to the typical interaction length of the system, 
  a series of universal three-body bound states can be found with a fixed factor between them. In the 
  unitary limit, $\frac{1}{a} = 0$, the energies of consecutive states fulfil
  \begin{equation}
    \frac{E_{n+1}}{E_n} = \lambda,
  \end{equation}
  where $\lambda$ is a universal factor that does not depend on details of the interaction. 
  It does depend however on the number of interacting pairs and the mass imbalance, among other 
  things. The typical spectrum resulting from this is usually depicted in a so-called Efimov plot
  as presented in Fig.~\ref{efimovplot}.

  The most easily accessible systems experimentally 
  are systems of two species of atoms (heavy bosons $H$ and one light atom $L$) with a large mass imbalance.
  There the Efimov effect occurs as well and the large mass imbalance 
  leads to a comparatively small factor $\lambda$ between consecutive states~\cite{hammerreview}. 
  
  In addition to the standard three-body effect, there has also been a lot of interest in 
  expanding the universal picture into the $N$-body sector. In systems of identical bosons, 
  for example, two universal tetramers attached to the lowest Efimov trimer have been 
  found~\cite{hammerplatter2007}\cite{stecherdincao2009}\cite{deltuva2012hpw}.
  
  Bringing these two aspects together, Blume and Yan have studied four-body states in a 
  heteronuclear system~\cite{blumeyan2014}. They focussed on the negative scattering length 
  side of the Efimov spectrum, because this is the side most experiments use for their observations. 
  
  To complete this picture we have studied the side of positive scattering length. This side can 
  also be seen experimentally, although experiments are complicated by the existence of the shallow dimer. 
  Experiments typically measure atomic losses due to resonant recombination at the threshold, so we have focussed on
  the region near the dimer threshold using full four-body calculations with the Gaussian Expansion 
  Method~\cite{hiyamakino2003}.

  \section{Method}
  \label{method}
   The Gaussian Expansion Method employs the Rayleigh-Ritz Variational Principle with 
   base functions that are selected via geometric progression between a minimum and a maximum range.
   It uses various representations of the system in Jacobi Coordinates, which we call rearrangement channels, to 
   cover different possible angular momentum couplings and arrangements. 
   The total wavefunction $\Psi$ the Schr\"odinger Equation $(H-E)\Psi = 0$
   is projected onto 
   is composed of a set of Gaussian functions for each Jacobi Coordinate, so in the three-body case we have
   \begin{equation}
    \Psi^{\rm trimer}_{JM} =  \sum^2_{c=1} \sum_{n_c,N_c}\sum_{\ell_c,L_c} C^{(c)}_{n_c \ell_c N_c L_c} 
    [\phi^{(c)}_{n_c \ell_c}({\bf r}_c) \psi^{(c)}_{N_c L_c}({\bf R}_c)]_{JM}, 
    \end{equation}
    with
    \begin{equation}
     \phi_{nlm}({\bf r})  = Y_{lm}({\widehat {\bf r}}) N_{nl}\,r^l\:e^{-(r/r_n)^2} \qquad\text{and}\qquad
     \psi_{NLM}({\bf R})  = Y_{LM}({\widehat {\bf R}}) N_{NL}\,R^L\:e^{-(R/R_N)^2}.
    \end{equation}
   \begin{figure}[b]
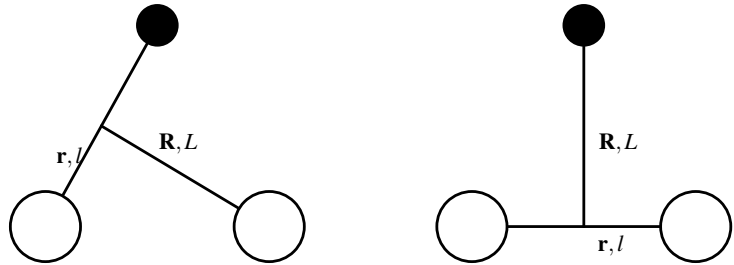

    \centering
   \begin{tikzpicture}[x=0.9cm,y=0.9cm]
      \node at (1.7,0.5) {\includegraphics[width=0.18\textwidth]{trimerchannelshetero.1}};
   \node at (8,0.5) {\includegraphics[width=0.18\textwidth]{trimerchannelshetero.2}};
   \node at (0.4, 0.2) {$\vec r, l$};
   \node at (2, 0.4) {$\vec R, L$};
   \node at (8.4, -1.1) {$\vec r, l$};
   \node at (8.5, 0.4) {$\vec R, L$};
   \end{tikzpicture}
   \caption{The two possible rearrangement channels for the $H_2L$ three-body system. The big white circles 
   represent the heavy bosons $H$, while the small black circle 
  represents the light atom $L$.}
   \label{jacobi}
   \end{figure}
   The relevant rearrangement channels for the trimer are shown in Fig.~\ref{jacobi}.\\
   The $r_n$ are given by 
   \begin{equation}
    r_n =r_{\rm min}\, \left(\frac{r_{\mathrm{max}}}{r_{\mathrm{min}}}\right)^\frac{n-1}{n_{\rm max}-1},\qquad n =1,..,n_{\rm max}\;,
   \end{equation}
   and analogously for $R_N$, which means for each Jacobi coordinate we have 3 parameters, $r_{\mathrm{max}}$, $r_{\mathrm{min}}$ and $n_{\mathrm{max}}$.
   More details about this method can be found in~\cite{hiyamakino2003}.
   Using several rearrangement channels and relative angular momentum configurations, the parameter space increases 
   rapidly for higher $N$ systems. Because of this, systematic sampling of base functions was impractical for the 
   four-body system and we employed random sampling within relatively broad ranges to find optimized base functions.
   
   \section{Interaction}
   Because there is no reason why the $HL$ interaction should be resonant at the same point as the $HH$ interaction, 
   we can safely assume $a_{HH} \ll a_{HL}$ and neglect $a_{HH}$. Adding an interaction between the heavy bosons only 
   moves the whole Efimov spectrum up or down~\cite{wanglaing2012}, which we can do independently by tuning our three-body force.
   
   As we want to focus on universal behaviour we choose effective potentials, which have the added benefit of 
   being numerically well behaved. To keep the calculated energies in the range of validity of the effective theory
   we add a repulsive three-body force. The inverse scattering length $\frac{1}{a}$ is roughly linear in the 
   two-body potential strength, which is 
   how we tune it.
        \begin{equation}
         V_{iN} = V_0 \exp\left({-\frac{r_{iN}^2}{2r_0^2}}\right), 
         \qquad W_{ijN} = W_0 \exp\left({-\frac{r_{ij}^2 + r_{jN}^2 + r_{iN}^2}{16r_0^2}}\right).
        \end{equation}
   Here $i$ and $j$ run over the heavy ($H$) bosons of mass $M$ from $1,..,N-1$ , while the $N$th atom ($L$) is light and of mass $m$.  
   The $r_{ij}$ are the distances between atoms $i$ and $j$.
  The natural energy scale of the problem is
  \begin{equation}
    E_s = \frac{1}{2r_0^2}\frac{m+M}{Mm},
  \end{equation}
  where $r_0$ is the effective range of the two-body potential. We have ensured $E \ll E_s$  throughout our calculations.  

  \section{Results}
  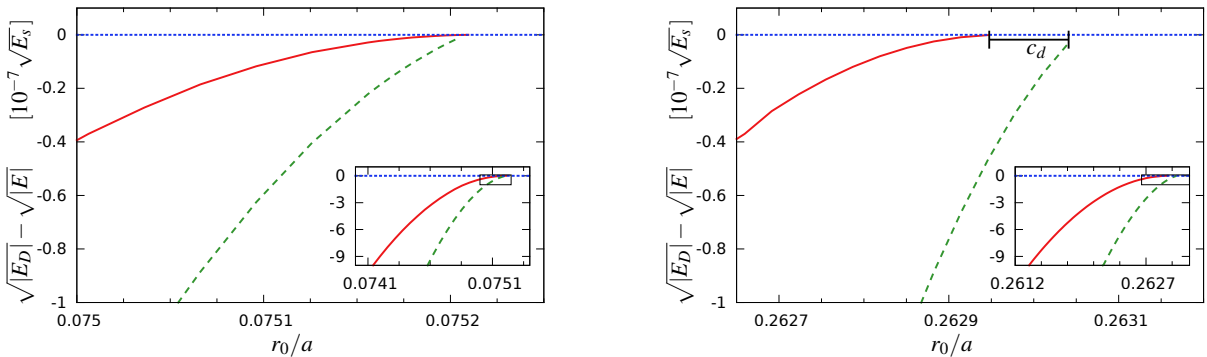
\begin{figure}[b]
  \vspace{-3ex}
   \begin{tikzpicture}[x=0.27cm,y=0.27cm] 
 \node at (1,0) {\includegraphics[width=0.45\textwidth]{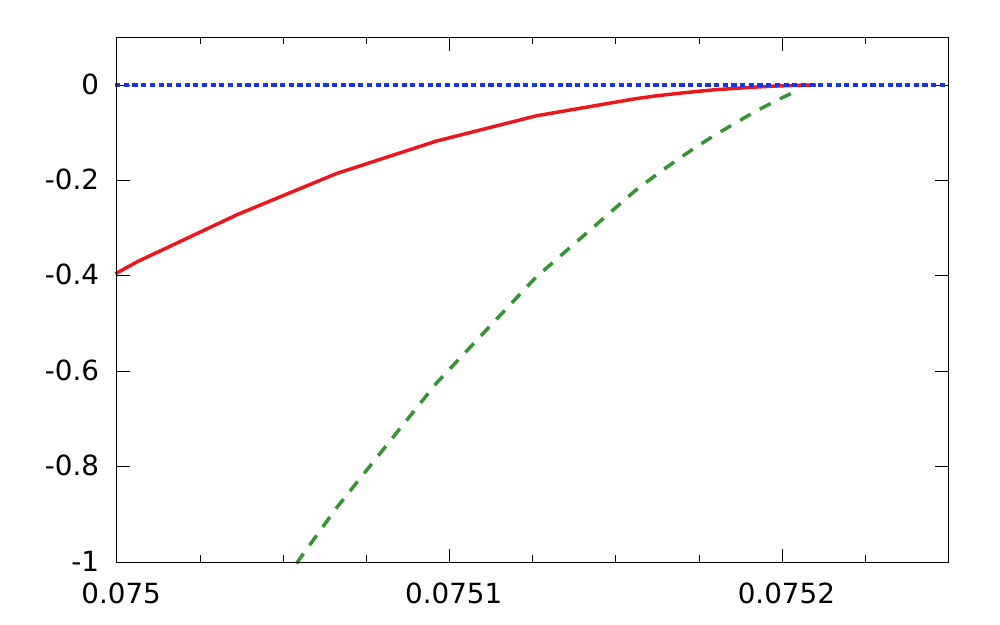}};
 \node[rotate=90, anchor=north] at (-13.5,0) {$\sqrt{\abs{E_D}}-\sqrt{\abs{E}}\qquad [10^{-7}\sqrt{E_s}\,]$};
 \node at (1,-9.){$r_0/a$};
 \node at (8,-3) {\includegraphics[width=0.2\textwidth]{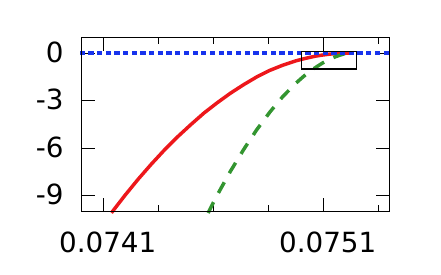}};
 \end{tikzpicture}\hfill
\begin{tikzpicture}[x=0.27cm,y=0.27cm]
 \node at (1,0) {\includegraphics[width=0.45\textwidth]{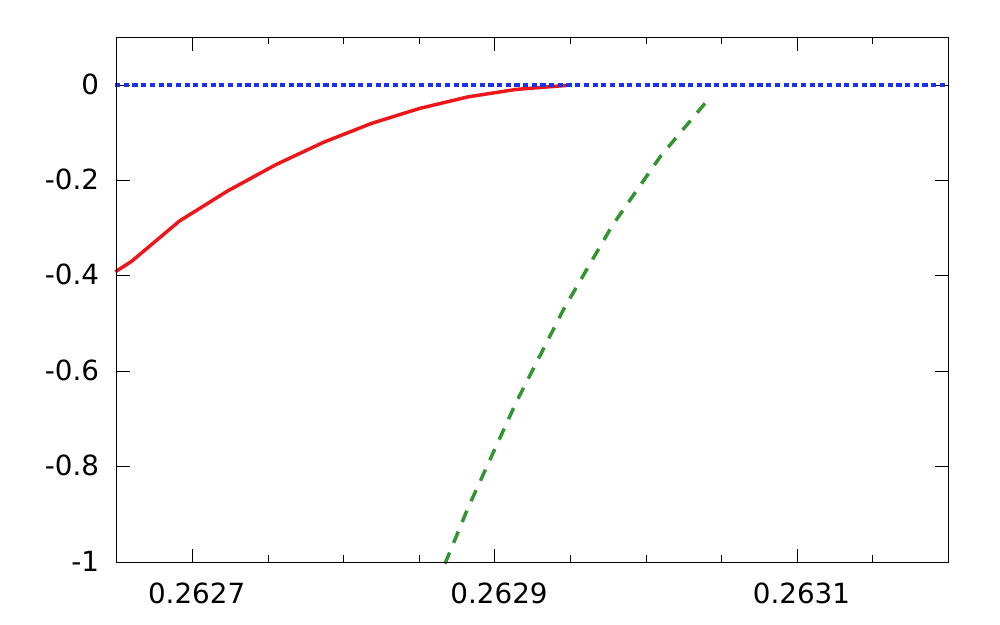}};
 \node[rotate=90, anchor=north] at (-13.5,0) {$\sqrt{\abs{E_D}}-\sqrt{\abs{E}}\qquad [10^{-7}\sqrt{E_s}\,]$};
 \node at (1,-9.){$r_0/a$};
 \node at (8,-3) {\includegraphics[width=0.2\textwidth]{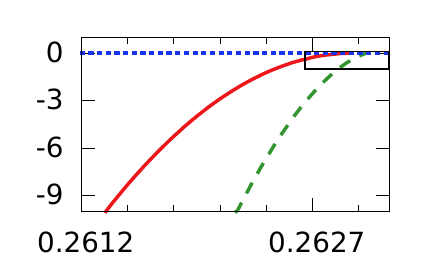}};
 \path[|-|](2.8,6.05) edge [line width=0.7]  (6.8,6.05);
 \node at (5.2,5.5) {$c_d$};
 \end{tikzpicture}
 \vspace{-1.7ex}
 \caption{Region of the Efimov plot very close to the dimer threshold. We display the difference to the dimer threshold (blue dotted line) 
 of the $H_2L$ trimer (red solid line) and the $H_3L$ tetramer (green dashed line) for two different systems, $^7$Li-$^6$Li mixture in the left panel
 and $^{133}$Cs-$^6$Li mixture in the right one. The insets are the zoomed out versions of the plots. We have included an illustration of the definition of $c_d$ 
 in the right plot.}
 \label{nearthreshold}
  \end{figure}

  Our main result is that trimer and tetramer vanish at almost the same point for all systems 
  we have looked at, which were $^7$Li-$^6$Li, $^{87}$Rb-$^7$Li and $^{133}$Cs-$^6$Li mixtures. 

\begin{figure}
  \centering
  \vspace{-3ex}
   \begin{tikzpicture}[x=0.3cm,y=0.3cm]
 \node at (1,0.3) {
 \includegraphics[width=0.5\textwidth]{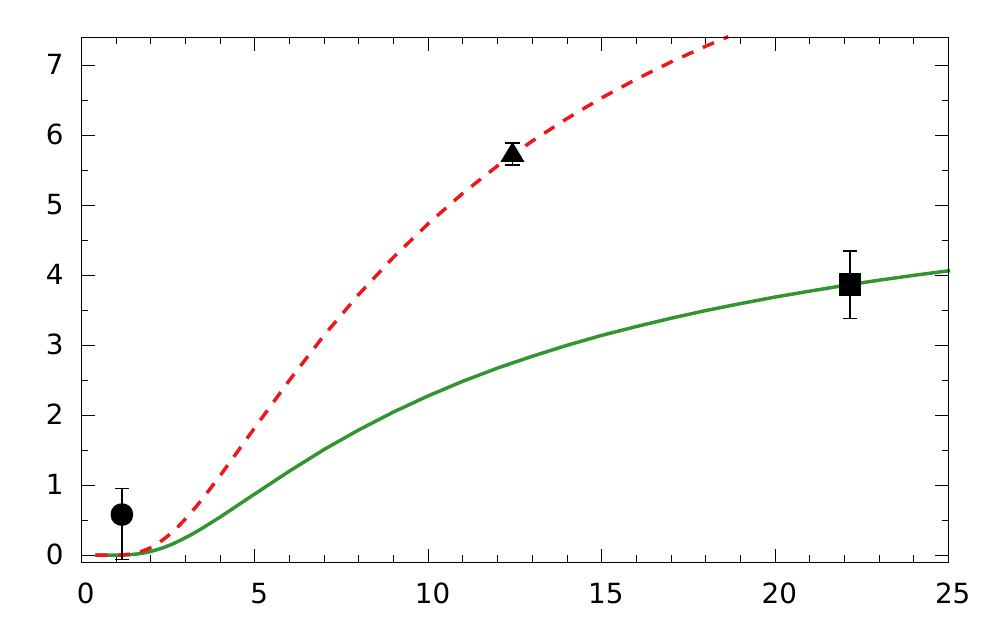}
 };
 \node[rotate=90, anchor=north] at (-13.5,0) {$c_d/c_3\times 10^{4}$};
 \node at (1.9,-8){$M/m$};
 \node at (-7.9,-3.4){$^7$Li$/^6$Li};
 \node at (-0.5,5.5){$^{87}$Rb$/^7$Li};
 \node at (10,-0.5){$^{133}$Cs$/^6$Li};
  \end{tikzpicture}
  \vspace{-1ex}
  \caption{The difference $c_d$ between trimer and tetramer crossing point in units of the trimer crossing point $c_3$ 
  as a function of the mass ratio $M/m$. The solid green (dashed red) lines are effective three-body calculations using 
  the Skorniakov-Ter-Martirosian equation
  and fitted to reproduce the $133/6$ ($87/7$) point.}
  \label{cd}
  \end{figure}
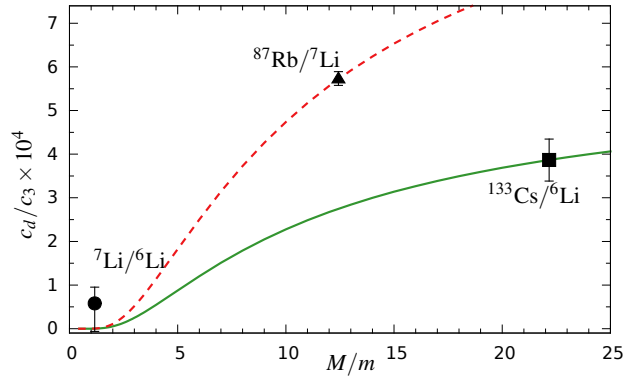

  We have taken a closer look at the behaviour of trimer and tetramer near the 
  dimer threshold. Our results for the two most extreme cases, almost no mass imbalance ($^7$Li-$^6$Li) 
  and very high mass imbalance ($^{133}$Cs-$^6$Li), are shown in Fig.~\ref{nearthreshold}. It can be seen 
  that in the (7/6) case the states approach the threshold very slowly, whereas for the (133/6) case 
  there is a slightly steeper slope. Also, the difference $c_d$ between the 
  point where the trimer vanishes, $c_3$, and the point where the tetramer vanishes, $c_4$, as 
  shown in the right panel of Fig.~\ref{nearthreshold}, is much larger for the (133/6) case than 
  for the (7/6) case. To quantify this, we have extracted these numbers for the three systems we have 
  investigated and plotted them in Fig.~\ref{cd}. 

  To help interpret this picture, we carried out effective three-body calculations of the dimer-atom-atom system 
  using the Skorniakov-Ter-Martirosian (STM) equation as described in \cite{helfrichhammer2010}. This picture is valid near the dimer threshold where the trimer 
  and tetramer are
  very shallow. In this region the scattering length between $HL$ dimer and $H$ atom can be estimated by the inverse trimer binding 
  energy and is very large ($a_{AD}/r_0 \approx 10^7$) so an effective three-body treatment is justified.
  The STM equation has one free parameter which we fitted to reproduce 
  either the data point belonging to the $(133/6)$-system, or the $(87/7)$ system. In both cases, the data point of the $(7/6)$-system 
  is reproduced, but the $(133/6)$ and $(87/7)$ points cannot be fitted simultaneously. This might mean that there is 
  non-universal behaviour governing the $c_d/c_3$ value.
  But it could also simply be the case that the effective three-body calculation does not capture 
  enough of the four-body system to yield accurate predictions.
   
  \section{Outlook}
  As a next step, we plan to investigate the dependence of $c_d$ on the 
  potential shape and the strength of the three-body potential to rule 
  out the possibility of non-universal behaviour. 
  Additionally, calculating more data points for Fig.~\ref{cd} might deepen 
  the understanding of the underlying functional dependence.
  
 \begin{acknowledgements}
 We thank Doerte Blume for useful discussions and comments. Part of the calculations were 
 carried out 
 at YITP of Kyoto University. C.H. Schmickler thanks RIKEN for providing support under 
 the IPA program.  
 \end{acknowledgements}
 


\end{document}